\documentclass[aip,nofootinbib,12pt]{revtex4-1}
\usepackage{amssymb,amsmath,amsthm}
\usepackage{textcomp}
\usepackage{latexsym}
\usepackage[english]{babel} 
\usepackage[latin1]{inputenc}  
\usepackage[T1]{fontenc}   
\usepackage[all]{xy}
\usepackage{amsfonts}
\usepackage{graphics}
\usepackage{epsfig}
\usepackage{graphicx}
\usepackage[all]{xy}
\usepackage[colorlinks=true,linkcolor=blue]{hyperref}%
\newcommand{\color}[6]{}

\def\CC{\mathbb{C}}

\def\C{\mathbb{C}}

\def\PP{\mathbb{P}}

\newcommand{\incl}{\ar@{^{}-}}

\newtheorem{prop}{Proposition}[section]

\newtheorem{lemma}{Lemma       }
\newtheorem{def }{Definition  }[section]
\newtheorem{algo}{Algorithm}[section]
\newtheorem{theorem}{Theorem}
\newtheorem{ex}{Example     }[section]
\newtheorem{rem}{Remark    }[section]

\begin{document}
\title{Singularity of type $D_4$  arising from four qubit systems}
\author{Fr\'ed\'eric Holweck\footnote{frederic.holweck@utbm.fr, 
 Laboratoire IRTES-M3M,
 Universit\'e de Technologie de Belfort-Montb\'eliard, 90010 Belfort Cedex, FR}, 
 Jean-Gabriel Luque\footnote{jean-gabriel.luque@univ-rouen.fr, Universit\'e de Rouen, Laboratoire d'Informatique, du Traitement de l'Information et des Syst\`emes (LITIS), Avenue de l'Universit\'e - BP 8
 6801 Saint-\'etienne-du-Rouvray Cedex, FR }
Michel Planat\footnote{michel.planat@femto-st.fr, Institut FEMTO-ST, CNRS, 32 Avenue de l'Observatoire, 25044 Besan\c{c}on, FR}}

\begin{abstract}
 An intriguing correspondence between four-qubit systems and simple singularity of type $D_4$ is established.
 We first consider the algebraic variety $X$ of separable states within the projective Hilbert space $\PP(\mathcal{H})=\PP^{15}$.
 Then, cutting $X$ with a specific hyperplane $H$, we prove that the $X$-hypersurface, defined from the section $X\cap H\subset X$, has an isolated 
 singularity of type $D_4$; it is also shown that this is the ``worst-possible'' isolated singularity one can obtain by this construction.
 Moreover, it is demonstrated that this correspondence admits a dual version by proving that the equation of the dual variety of $X$, which is nothing but the Cayley 
 hyperdeterminant of type $2\times 2\times 2\times 2$, can be expressed in terms of the SLOCC invariant polynomials as the discriminant of the miniversal deformation of the $D_4$-singularity.
 \end{abstract}


\maketitle

\section{Introduction}

Several branches of geometry and algebra tend to play an increasing role in quantum information theory. 
We have in mind algebraic geometry for describing entanglement classes of multiple 
qubits\cite{brody1,My2,hey,HLT}, representation theory and Jordan algebras for entanglement and the black-hole/qubit correspondence\cite{BDDER,BDDER2,BDFMR}, and geometries over finite fields/rings for deriving point-line configurations of observables relevant to quantum contextuality\cite{LSV,SLPP,PSH}.  
The topology of hypersurface singularitites, and the related Coxeter-Dynkin diagrams, represent another field worthwhile to be investigated in quantum information, as shown in this paper.


Dynkin diagrams are well known  for classifying simple Lie algebras, Weyl groups, 
subgroups of $SU(2)$ and simple singularities, \emph{i.e.} isolated singularities of complex hypersurfaces that are stable under small perturbations.
More precisely, if we consider simple-laced Dynkin diagrams, \emph{i.e.} diagrams of type $A-D-E$, we find objects of different nature 
classified by the same diagrams:
\[\begin{array}{c|c|c|c}
 \text{Type}  & \text{Lie algebra} & \text{Subgroup of } SU(2) & \text{Hypersurface with simple singularity} \\
   \hline
   A_n & \mathfrak{s}\mathfrak{l}_{n+1}(\CC) & \text{cyclic group} & x_1^{n+1}+x_2^2+\dots+x_k^2=0\\
   D_n & \mathfrak{s}\mathfrak{o}_{2n} (\CC) & \text{binary dihedral group} & x_1^{n-1} +x_1x_2^2+x_3^2+\dots+x_k^2=0\\
   E_6 &\mathfrak{e}_6 &\text{binary tetrahedral}  & x_1^4+x_2^3+x_3^2+\dots+x_k^2=0\\
   E_7 &\mathfrak{e}_7 &\text{binary octahedral} & x_1^3x_2+x_2^3+x_3^2+\dots+x_k^2=0\\
   E_8  &\mathfrak{e}_8 &\text{binary icosahedral} & x_1 ^5+x_2^3+x_3^2+\dots+x_k^2=0
  \end{array}\]
A challenging question in mathematics is to understand these $ADE$-correspondences by establishing a direct construction 
from one class of objects to the other.
For instance, the construction of surfaces with simple singularities from the corresponding subgroup of $SU(2)$ is called the McKay correspondence.
A construction due to Grothendieck allows us to recover the simple singularities of a given type from the nullcone (the set of nilpotent elements) of the corresponding 
simple Lie algebra. For an overview of such $ADE$ correspondences, see Ref\cite{slodowy1,slodowy2} and references therein.

Another construction connecting simple Lie algebras and simple singularities is due to Knop\cite{Kn}. 
In his construction, Knop considers a unique smooth
orbit, $X$, for the adjoint action of Lie group $G$ on the projectivization of its Lie algebra $\PP(\mathfrak{g})$ and cuts this 
variety by a specific hyperplane. The resulting 
$X$-hypersurface has a unique singular point of the same type as $\mathfrak{g}$.

Looking at $ADE$-correspondences in the context of QIT is a way to understand the role played by those diagrams in this field. In different classification schemes of four-qubit systems, the Dynkin diagram $D_4$ has already appeared thanks to the role played by the Lie algebra 
$\mathfrak{so}(8)$ (that is the type $D_4$). For instance, Verstraete {\em et al}'s classification\cite{VDMV} is based on the classification of  the  
$SO(4)\times SO(4)\subset SO(8)$ orbits on $\mathcal{M}_4(\CC)$. Chterental and Djokovic\cite{CD} use the same
group action and refer to (Remark 5.3 of Ref\cite{CD}) the Hilbert space of four qubits as a 
subspace of $\mathfrak{so}(8)$ whose SLOCC orbits arise from the trace of the adjoint $SO(8)$ orbits. In their study of the four-qubit 
classification from the string theory point of view, 
Borsten {\em et al}\cite{BDD} employ a correspondence between nilpotent orbits of $\mathfrak{so}(4,4)$ (the real form of $\mathfrak{so}(8)$ with signature $(4,4)$) and nilpotent orbits of four-qubit systems.
Last but not least, the relation between $\mathfrak{so}(8)$ and four-qubit systems has been pointed out by L\'evay\cite{Levay1} in his paper on the black-hole/qubit correspondence. In this paper L\'evay describes the Hilbert space of four qubits as the tangent space 
of $SO(4,4)/(SO(4)\times SO(4))$.

In the present paper, we will establish a correspondence between four-qubit systems and $D_4$-singularities by using a construction
inspired by Knop's paper. In other words, we will establish an $ADE$-type correspondence between $SO(4,4)$ and singularities of type $D_4$ using the Hilbert space of four qubits. 

Let $\mathcal{H}=\CC^2\otimes\CC^2\otimes\CC^2\otimes \CC^2$ be the Hilbert space
of four-qubit systems. Up to scalar multiplication, a four-qubit $|\Psi\rangle\in \mathcal{H}$ can be considered as a point of the 
projective space $\PP^{15}=\PP(\mathcal{H})$. The set of separable states in $\mathcal{H}$ corresponds to tensors of rank one, \emph{i.e.}
tensors which can be factorized as $|\Psi\rangle=v_1\otimes v_2\otimes v_3\otimes v_4$ with $v_i\in \CC^2$. Adopting the notation 
$\{|0\rangle,|1\rangle\}$ for the single-qubit computational basis and $|ijkl\rangle=|i\rangle\otimes|j\rangle\otimes |k\rangle\otimes |l\rangle$ for the four-qubit basis, a general four-qubit state can be expressed as 

 \[|\Psi\rangle=\sum_{0\leq i,j,k,l\leq 1} a_{ijkl}|ijkl\rangle \text{ with } a_{ijkl}\in \CC.\]
Let $G$ be the 
group of  Stochastic Local 
Operation and Classical Communication (SLOCC) of four qubits [acting on $\PP(\mathcal{H})$], \emph{i.e.} 
$G=SL_2(\CC)\times SL_2(\CC)\times SL_2(\CC)\times SL_2(\CC)$. It is well known that $G$ acts transitively 
on the set of separable states. The projectivization of the corresponding orbit -- also called the highest weight orbit --
is the unique smooth orbit $X$ for the action of $G$ on $\PP(\mathcal{H})$, that is
\[X=\PP(G.|0000\rangle)=\{\text{The set of separable states}\}\subset \PP^{15}.\]

A parametrization of $X$ is given by the Segre embedding of four projective lines\cite{hey,HLT}
\[\phi:\left\{\begin{array}{ccc}
       \PP^1\times\PP^1\times\PP^1\times\PP^1 & \to & \PP^{15} \\
       ([w_0:w_1],[x_0:x_1],[y_0:y_1],[z_0:z_1]) & \mapsto &[w_0x_0y_0z_0:\dots:W_J:\dots:w_1x_1y_1z_1]
       \end{array}\right.\]where $W_J=w_ix_jy_kz_l$ for $J=\{i,j,k,l\}\in\{0,1\}^4$ and the monomial order is such 
       that $W_{J_1}\prec W_{J_2}$ if $8i_1+4j_1+2k_1+l_1\leq 8i_2+4j_2+2k_2+l_2$.

A hyperplane $H\subset \PP(\mathcal{H})$ is the set of states $|\Phi\rangle\in \PP(\mathcal{H})$ on which a linear 
form $L_{H}\in \mathcal{H}^*$ vanishes. Given $H\subset \PP(\mathcal{H})$, the hyperplane section $X\cap H\subset X$ is the hypersurface 
of $X$ defined by the restriction of $L_H$ to $X$.  Due to the 
duality of Hilbert spaces, for any $H\subset \PP(\mathcal{H})$
there exists a state $|\Psi\rangle \in \PP(\mathcal{H})$ such that 
$H$ is defined by the linear form $\langle \Psi|$. In what follows, we will often 
identify the hyperplane $H$ and the linear form defining it, and write $H=\langle \Psi|=\sum_{0\leq i,j,k,l\leq 1} h_{ijkl}\langle ijkl|$ with $h_{ijkl}\in \C$. The hyperplane section $X\cap H$, or, equivalently, 
$X\cap \langle \Psi|$, will be the hypersurface of $X$ given by 
\begin{equation}\label{section_eq}
 \langle \Psi|\phi(\PP^1\times\PP^1\times\PP^1\times\PP^1)\rangle=\sum_{0\leq i,j,k\leq 1}h_{ijkl} w_ix_jy_kz_l=0.
\end{equation}

To state our main Theorem, let us recall that the ring of polynomials invariant under $G$ is generated by $4$ invariants\cite{LT}. Let us denote by $\tilde{I}_1,\tilde{I}_2,\tilde{I}_3,\tilde{I}_4$
a choice of four generators of the ring of invariants (that choice will be explained in Section \ref{hyperdeterminant}),  \emph{i.e.} 
$\CC[\mathcal{H}]^G=\CC[\tilde{I}_1,\tilde{I}_2,\tilde{I_3},\tilde{I}_4]$. The quotient map $\Phi:\mathcal{H}\to \CC^4$ is defined by 
$\Phi(x)=(\tilde{I}_1(x),\tilde{I}_2(x),\tilde{I}_3(x),\tilde{I}_4(x))$.
The main result of this article is the following theorem:
\begin{theorem}\label{mainthm}
 Let $H=\langle \Psi|$ be a hyperplane of $\PP(\mathcal{H})$ tangent to $X$ and such that $X\cap H$ has 
 only isolated singular points. Then 
 the singularities are either of types  $A_1, A_2, A_3, A_4$, or of type $D_4$, and there exist hyperplanes realizing each type of singularity.
 Moreover, if we denote by $\widehat{X}^*\subset \mathcal{H}$ the cone over the dual variety of $X$, \emph{i.e.} the zero locus of the Cayley
 hyperdeterminant of format $2\times 2\times 2\times 2$, then the quotient map $\Phi:\mathcal{H}\to \CC^4$ is such that $\Phi(\widehat{X}^*)=\Sigma_{D_4}$, where
 $\Sigma_{D_4}$ is the discriminant of the miniversal deformation of the $D_4$-singularity.
\end{theorem}
The paper is organized as follows. In Section \ref{hyperplane_section}, we will give the definition of a simple singularity
and the invariants that follow from the Arnol'd classification\cite{A} (Section \ref{Arnold}). Then we will compute the singularity type of any 
hyperplane section of the set of separable states featuring only  isolated singularities (see Section \ref{calcul_section} Proposition \ref{section}).
In Section \ref{caley}, we will establish a dual version of Proposition \ref{section}. We will first define the notion of discriminant of a singularity (see Section \ref{disc})
and then show how it allows us to give a new expression for the Cayley hyperdeterminant $\Delta_4$ (Section \ref{hyperdeterminant}) and  prove Proposition \ref{dual} about the relation between $\Delta_4$ and $\Sigma_{D_4}$. Propositions \ref{section} and \ref{dual} lead to the proof of  Theorem \ref{mainthm}.
\section{Simple singularities and hyperplane sections of separable states}\label{hyperplane_section}

\subsection{Simple singularities following Arnol'd classification}\label{Arnold}
Simple singularities  have been studied from an algebraic geometrical viewpoint  
as rational double points of algebraic surfaces, Du Val singularities, and from a complex 
analytic perspective as critical points of holomorphic functions in
several variables. These approaches lead to many equivalent characterizations of what a simple 
singularity is\cite{Du}. Here, we select the complex analytic approach introduced by Vladimir Arnol'd.
We first recall the ingredients of Arnol'd classification of simple singularities \cite{A}.

Let us denote by $(f,0)$ the germ of a holomorphic function, $f:(\CC^k,0)\to (\C,0)$ at $0$, and by $\mathcal{O}_k$ 
the set of all those germs. We consider the group $\mathcal{D}_k$ of biholomorphic maps $g:(\CC^k,0)\to (\CC^k,0)$ acting of $\mathcal{O}_k$
such that $g.f=f\circ g^{-1}$. A {\em singularity} is an equivalence class of a germ $(f,0)$ such that 
$\dfrac{\partial f}{\partial x_i}(0)=0$ for $i=1,\dots,k$.
In other words, a singularity is an orbit in $\mathcal{O}_k$ and we will write $[(f,0)]$ for the orbit of the representative $(f,0)$. We denote by $\mathcal{S}_k\subset \mathcal{O}_k$ the set of all singular germs.
Let $f$ be a representative of a singularity and let us denote by $A=\left(\dfrac{\partial^2 f}{\partial x_i \partial x_j}(0)\right)_{i,j}$
the corresponding Hessian matrix. The corank of the germ $(f,0)$ is the dimension of the kernel of $A$. 
From the definition of the action of $\mathcal{D}_k$ it follows that equivalent germs will have the same corank, which means that the corank is
an invariant of a singularity.
\begin{def }
 A singularity is said to be non-degenerate, or quadratic, or of the Morse type, if, and only if, 
 its corank is zero.
\end{def }

The Morse Lemma \cite{Mi} ensures that if $(f,0)$ is a non-degenerate singular germ, 
then $f\sim x_1^2+\dots +x_k^2$. The non-degenerate singularity is a dense orbit in $\mathcal{S}_k$.
Assume that $[(f,0)]$ is a singularity of corank $l$, a generalization of Morse's Lemma\cite{A} 
tells us that $f\sim h(x_1,\dots,x_l)+x_{l+1}^2+\dots+x_k^2$ and leads to an equivalence relation between germs of distinct number of variables.

\begin{def }
 Two function germs $f:(\CC^k,0)\to (\CC,0)$ and $g:(\CC^m,0)\to (\CC,0)$, with $k<m$, are said to be stably equivalent
 if, and only if, $f(x_1,\dots,x_k)+x_{k+1}^2+\dots+x_m^2\sim g(x_1,\dots,x_m)$.
\end{def }

\begin{rem}\rm
 In terms of the last definition we can compare singularities of functions which do not have the same number of variables. Adding quadratic terms of full rank
 in new variables do not affect the classification of the singular type.
\end{rem}

Another important invariant of singular germs is the famous Milnor number\cite{Mi}. Let $(f,0)$ be a singular germ and consider 
$I_{\nabla f}=\mathcal{O}_k<\dfrac{\partial f}{\partial x_1}(0),\dots,\dfrac{\partial f}{\partial x_k}(0)>$  the gradient ideal.

\begin{def }
 The Milnor number $\mu$ of a singular germ $(f,0)$ is equal to the dimension of the local algebra of $(f,0)$, \emph{i.e.} the quotient of 
 the algebra $\mathcal{O}_k$ by $I_{\nabla f}$,
 \[\mu=\text{dim}_\CC \left( \mathcal{O}_k\diagup I_{\nabla f}\right).\]
\end{def }

The critical point $0$ of the function $f$ will be isolated if, and only if, its Milnor number is finite.

Let us now state what, in the sense of Vladimir Arnol'd, a simple singularity is .

\begin{def }\label{simple}
 The orbit $[(f,0)]$ is a simple singularity if, and only if,  a sufficiently small neighborhood of $(f,0)$ intersects $\mathcal{S}_k$
 with a finite number of non-equivalent orbits. 
\end{def }

\begin{rem}\rm
 If we consider a representative of a non-degenerate singularity $f\sim x_1^2+\dots+x_k^2$, a small perturbation of  
 $f$ in $\mathcal{S}_k$, \emph{i.e.} $f+\varepsilon h$ with 
 $h\in \mathcal{S}_k$, will still have a Hessian of full rank for $\epsilon$ small. Thus $f\sim f+\varepsilon h$, which means that non-degenerate
 singularity is the most stable type of singularity. We can rephrase Definition \ref{simple} by saying that $[(f,0)]$
 is a simple singularity if, and only if, a small perturbation of a representative $f$ will only lead to a finite number of non-equivalent singularities.
\end{rem}

In his classification of simple singularities\cite{A}, Arnol'd proved that being simple is equivalent to the following conditions:
\begin{itemize}
\item $\mu<+\infty$,
 \item $\text{corank}\left(\dfrac{\partial^2 f}{\partial x_i \partial x_j}(0)\right)\leq 2$,
 \item if $\text{corank}\left(\dfrac{\partial^2 f}{\partial x_i \partial x_j}(0)\right)=2$ the cubic term in the degenerate direction of the Hessian is non-zero,
 \item if $\text{corank}=2$ and the cubic term is a cube then $\mu<9$.
\end{itemize}

With these conditions Arnol'd obtained the classification of simple singularities into five different types (Table \ref{simple_sing}).

\begin{table}[!h]
 \begin{tabular}{c|c|c|c|c|c}
  Type & $A_n$ & $D_n$ & $E_6$ & $E_7$ & $E_8$ \\
  \hline
  Normal forms & $x^{n+1}$ & $x^{n-1}+xy^2$ & $x^3+y^4$ & $x^3+xy^3$ & $x^3+y^5$ \\
  \hline
  Milnor number & $n$ & $n$ & $6$ & $7$ & $8$ \\
 \end{tabular}
\caption{Simple singularities.}\label{simple_sing}
\end{table}

\begin{rem}\rm
 The functions given in Table \ref{simple_sing} are stably equivalent to the hypersurfaces given in the introduction.
 They are also clearly equivalent to the rational double points of algebraic surfaces.
\end{rem}

The classification given by Arnol'd furnishes an algorithm to test if a singularity is simple or not.
\begin{algo}\label{algo} 
Let $(f,0)$ be a singularity.
 \begin{itemize}
  \item Compute $\mu$; if $\mu=\infty$ the singularity is not isolated (and not simple),
  \item If not, compute $r=\mbox{corank}(\mbox{Hess}(f,0))$.
  \begin{itemize}
  \item if $r\geq 3$, the singularity is not simple,
  \item if $r=1$, the singularity is of type $A_\mu$,
  \item if $r=2$, then 
  \begin{itemize}
  \item if the cubic term in the degenerate directions is non-zero and is not a cube, then the singularity is of type $D_\mu$,
  \item if the cubic term in the degenerate directions is a cube and $\mu <9$, then the singularity is of type $E_\mu$,
  \item if not, the singularity is not simple.
  \end{itemize}
  \end{itemize}
 \end{itemize}
\end{algo}

In the next section we will follow this algorithm to compute the singular type of a given hyperplane section.

\subsection{Computing singularities of hyperplane sections}\label{calcul_section}
Before we prove the first proposition, let us consider two examples in order to explain how we calculate the singular type of a hyperplane section.

\begin{ex}\rm\label{exsectionA1}
 Let $H\in \PP(\mathcal{H}^*)$ be a hyperplane, or a linear form, given by $H=\langle\Psi_1|=\langle 0011|+\langle 1100|$. The corresponding
 hyperplane section $X\cap H$ is tangent to $|1111\rangle$. Indeed, a tangent vector to $X$ at $|1111\rangle$ will be of the form 
 $|v\rangle=\alpha|0111\rangle+\beta|1011\rangle+\gamma|1101\rangle+\delta|1110\rangle$ and it is clear that $\langle \Psi_1|v\rangle=0$.
 The homogeneous form of the linear section $X\cap H$ corresponds to its restriction to (the cone over) $X$, that 
 is to \[f(w_0,w_1,x_0,x_1,y_0,y_1,z_0,z_1)=w_0x_0y_1z_1+w_1x_1y_0z_0.\]
 In a non-homogeneous form $f$ can be written in the chart corresponding to $w_1,x_1,y_1,z_1=1$ as
 $f(w_0,x_0,y_1,z_1)=w_0x_0+y_0z_0$. In this chart the point $|1111\rangle$ has coordinates $(0,0,0,0)$ and (we can forget about the subscripts)
 the hyperplane section is a hypersurface of $X$ defined (locally) by the equation \[f(w,x,y,z)=wx+yz=0.\]
 This hypersurface has a unique singularity $\left(\dfrac{\partial f}{\partial w}(a),\dfrac{\partial f}{\partial x}(a),\dfrac{\partial f}{\partial y}(a),\dfrac{\partial f}{\partial z}(a)\right)=(0,0,0,0)\Leftrightarrow a=(0,0,0,0)$,
which corresponds to $|1111\rangle$, and the Hessian matrix \[\begin{pmatrix}
                                                 0 & 1 & 0 & 0\\
                                                 1 & 0 & 0 & 0\\
                                                 0 & 0 & 0 &1\\
                                                 0 & 0 & 1 & 0
                                                \end{pmatrix}\]
                                                is of the full rank.
                                                One concludes that $(X\cap H,|1111\rangle)$ is an isolated singularity of type $A_1$ and
                                                we denote it by $(X\cap H,|1111\rangle)\sim A_1$, or, equivalently, by $(X\cap \langle \Psi_1|,|1111\rangle)\sim A_1$.

 \end{ex}

\begin{ex}\rm\label{exsectionD4}
 Let us consider the hyperplane section defined by $H=\langle \Psi_2|=\langle0000|+\langle1011|+\langle1101|+\langle1110|\in \mathcal{H}^*$.
 This section $X\cap H$ is tangent to $|0111\rangle$. It is clear that a tangent vector to $X$ at $|0111\rangle$ will be of the form 
 $|v\rangle=\alpha |1111\rangle+\beta|0011\rangle+\gamma|0101\rangle+|0110\rangle$ and $H|v\rangle=0$. The homogeneous linear
 form corresponding to 
 $X\cap H$ is $f(w_0,w_1,x_0,x_1,y_0,y_1,z_0,z_1)=w_0x_0y_0z_0+w_1x_0y_1z_1+w_1x_1y_0z_1+w_1x_1y_1z_0$.  In the chart $w_0=x_1=y_1=z_1=1$ 
 the form becomes a hypersurface defined by 
 \[xyz+wx+wy+wz=0\]
 and $(0,0,0,0)$ is the only singularity of this hypersurface.
 Using the software 
 SINGULAR\cite{DGPSS} one can check that $\mu_{x=(0,0,0,0)}(f)=4$ and the rank of the Hessian
 \[\begin{pmatrix}
0 & 1 & 1 & 1\\
1 & 0 & 0 & 0\\
1 & 0 & 0 & 0\\
1 & 0 & 0 & 0
                                                                                         \end{pmatrix}\]
is $2$. Thus, we conclude that $(X\cap H,|0111\rangle)\sim D_4$, or, equivalently, $(X\cap \langle \Psi_2|,|0111\rangle)\sim D_4$ 
(\emph{i.e.} the unique isolated singularity where the corank equals $2$ and $\mu=4$).
\end{ex}

We can now prove our first proposition.

\begin{prop}\label{section}
Let $X\cap H$ be a singular hyperplane section of the variety of separable states for four-qubit systems, 
\emph{i.e.} $X=\PP^1\times\PP^1\times\PP^1\times\PP^1$, with an isolated singularity $x\in X\cap H$. 
Then the singularity $(X\cap H,x)$ will be of type $A_1, A_2, A_3$ or $D_4$ and each type can be obtained by such a linear section of $X$.
\end{prop}


\proof To prove Proposition \ref{section}, we compute the singular type of all possible hyperplane sections of $X$. As the variety $X$ is $G$-homogeneous, the singular type of $X\cap H$ will be 
identical for any representative of the $G$ orbit of $H$. By the duality of the Hilbert space, 
a hyperplane $H$ corresponds to a point $h\in \PP(\mathcal{H})$. But the $G$ orbits of 
$\PP(\mathcal{H})$ have been classified by Verstraete {\em et al.}\cite{VDMV} 
(with a corrected version provided by  Chterental and Djokovic\cite{CD}).
According to Verstraete {\it et al.}'s classification, the $G$-orbits of the four-qubit Hilbert space 
consist of $9$ families ($3$ families are parameter free and $6$ of them depend on parameters) and normal 
forms for each family are known\cite{VDMV,CD}.
From each of Verstraete {\it et al.}'s normal forms $|\Psi\rangle$ we compute the 
corresponding hyperplane section $X\cap \langle \Psi|$. Then we look at isolated singular points of each hyperplane section and
we calculate the corresponding singular type with a formal
algebra system following the procedure described in examples \ref{exsectionA1}, \ref{exsectionD4} and Algorithm \ref{algo}. 
For the normal forms depending on parameters, 
the singular type of the hyperplane sections will depend on values of the parameters.
The results of our calculations are given in Tables \ref{form1} and \ref{form2} and provide a proof of the proposition.$\Box$
\begin{center}
\begin{table}[!h]
\begin{tabular}{|c|c|c|c|}
\hline
Verstraete {\it et al.}'s notation &  Hyperplane & Singular type of the hyperplane section \\
\hline
          $L_{0_{7\oplus \overline{1}}}$            &   $\langle0000|+\langle1011|+\langle1101|+\langle1110|$ & $D_4$ (a unique singularity )\\           
                      \hline
         $L_{0_{5\oplus \overline{3}}}$             &    $\langle0000|+\langle0101|+\langle1000|+\langle1110|$                     & non-isolated\\
                      \hline
    $L_{0_{3\oplus \overline{1}}0_{3\oplus \overline{1}}}$                  &       $\langle 0000|+\langle 0111|$                & non-isolated\\
                      \hline
 
\end{tabular}
\caption{Hyperplanes and the corresponding sections which do not depend on parameters.}\label{form1}
\end{table}
\end{center}
\begin{center}
\begin{table}[!h]
\begin{tabular}{|c|c|c|c|c|}
\hline
Verstraete's &  Hyperplane &  parameters & Singular type  \\
notation & & & \\
\hline
          $L_{a_20_{3\oplus {1}}}$            &   $a(\langle0000|+\langle1111|)+\langle0011|+\langle0101|+\langle0110|$ & $a$ generic & $A_1$\\           
                                              &                                                                     &$a=0$ &non-isolated \\
                                              \hline
                                  $L_{a_4}$ &    $a(\langle0000|+\langle0101|+\langle1010|+\langle1111|)$ &$a$ generic & $A_3$ (a unique singularity) \\
                                          & $+i\langle0001|+\langle0110|-i\langle1011|$ & $a=0$  & non-isolated\\
                                  \hline
                          $L_{ab_3}$ & $a(\langle0000|+\langle1111|)+\frac{a+b}{2}(  \langle0101|+\langle1010|)$ &$a,b$ generic &$A_2$ (a unique singularity)\\
                          &          $+\frac{a-b}{2}(  \langle0110|+\langle1001|)$ &$a=b=0$ & non-isolated \\
                          & $+\frac{i}{\sqrt{2}}(\langle0001|+\langle0010|-\langle0111|-\langle1011|)$ & &\\
                          
                      \hline
                      $L_{a_2b_2}$ & $a(|0000\rangle+|1111\rangle)+b(|0101\rangle+|1010\rangle)$& $a,b$ generic & smooth section \\
                                   &$+|0110\rangle+|0011\rangle$  & $a=0$ or $b=0$ & non-isolated\\
                                   &                              & $a=b=0$ & non-isolated\\
                                   \hline
 $L_{abc_2}$ &$\frac{a+b}{2}(\langle0000|+\langle1111|)+\frac{a-b}{2}(\langle0011|+\langle1100|)$ & $a,b,c$ generic & $A_1$ (a unique singularity)\\
& $c(\langle1010|+\langle0101|)+\langle0110|$  &  $a=\pm b$   & $A_1$ \\
 &                                   & $c=0$ & $A_1$  \\ 
 & & $a=\pm b=\pm c$ & non-isolated\\
 & & $a=c=0$ or $b=c=0$ & non-isolated\\
 &  & $a=b=c=0$ & non-isolated\\
       \hline
       $G_{abcd}$ &$\frac{a+d}{2}(|0000\rangle+|1111\rangle)+\frac{a-d}{2}(|0011\rangle+|1100\rangle)$   & $a,b,c,d$ generic & smooth section \\
       &  $+\frac{b+c}{2}(|0101\rangle+|1010\rangle)+\frac{b-c}{2}(|0110\rangle+|1001\rangle)$         & see Table \ref{Gabcd} & $A_1$ \\
         &         & see Table \ref{Gabcd_nonisole} & non-isolated \\
                  \hline
      
\end{tabular}
\caption{Hyperplanes and the corresponding sections which do  depend on parameters.}\label{form2}
\end{table}
\end{center}

\begin{rem}\rm
 Tables \ref{form2}, \ref{Gabcd}, \ref{Gabcd_nonisole} show that the classification of entangled states into 9 families can 
 be refined according to the singular type of the corresponding section. 
 The singular type of the linear section $X\cap \langle \Psi|$ is an invariant of the $G$-orbit 
 of $|\Psi\rangle$ and may be used to distinguish two non-equivalent classes of entanglement. 
 Thus, the values of the parameters which distinguish the sections indicate how we can decompose further the classification.
 However, to fully distinguish non-equivalent sections from their singular type, it would be necessary to investigate more precisely the non-isolated singular sections.
\end{rem}

\begin{rem}\rm
 It is worthwile to point out that the different isolated singular types we obtain by this 
 construction ($A_1,A_2,A_3$ and $D_4$) are exactly the possible
 degenerations of the $D_4$-singularity. In particular, any small neighborhood of the singularity of type $D_4$ will meet, in $\mathcal{S}_k$, the orbits  corresponding to the 
 singular types $A_1,A_2$ and $ A_3$ as shown in the adjacency diagrams of Arnold's classification (Corollary 8.7 in Ref\cite{A}). 
 The fact that $D_4$ is the ``worst-possible'' isolated singularity we get 
 from the hyperplane sections of the set of separable states will be lighted with Proposition \ref{dual}.
\end{rem}

\section{The Cayley $2\times 2\times 2\times 2$ hyperdeterminant and the $D_4$-discriminant}\label{caley}


Another fundamental concept associated with a simple singularity is its discriminant, \emph{i.e.} the locus 
that parametrizes the deformation of the singular germs. In this section, we will show that the discriminant of 
the $D_4$-singularity is linked to the dual variety, in the sense of the projective duality, of the set of separable four-qubit states.

\subsection{Discriminant of the miniversal deformation of the singularity}\label{disc}
Consider a holomorphic germ $f:(\CC^k,0)\to (\CC,0)$ with a simple isolated
singularity of Milnor number $\mu(f,0)=n$. A {\em miniversal deformation\cite{A}} of the germ $f$ is 
given by
\[f+\sum \lambda_i g_i,\]

where $(g_1,\dots,g_n)$ is a basis of $\mathcal{O}_k\diagup I_{\nabla f}$.
\begin{def }
 The discriminant $\Sigma\subset \CC^n$ is the subset of values $(\lambda_1,\dots,\lambda_n)\in \CC^n$ 
 such that the miniversal deformation 
 $f+\sum \lambda_i g_i$ is singular, \emph{i.e.}
 \[\Sigma=\{(\lambda_1,\dots,\lambda_n)\in \CC^n, \Delta(f+\sum_{i=1}^n \lambda_i g_i)=0\},\]
 where $\Delta$ is the usual notion of discriminant.
\end{def }

\begin{rem}\rm
 The discriminant parametrizes all singular deformations of $(f,0)$. 
 It is known\cite{Wir} that for hypersurfaces endowed with a simple singularity, the discriminant of the singularity 
 characterizes its type. 
\end{rem}

\begin{ex}\rm
 Let $(f,0)$ be a singularity of type $A_n$, \emph{i.e.} $f\sim x^{n+1}$. Then $\mathcal{O}_1\diagup I_{\nabla x^{n+1}}=<1,x,\dots,x^{n-1}>$. Thus, a miniversal deformation of $f$ is \[F(x,\lambda)=x^{n+1}+\lambda_1x^{n-1}+\lambda_2x^{n-2}+\dots+\lambda_n.\]
 The corresponding discriminant is the hypersurface $\Sigma_{A_{n}}\subset \CC^n$ defined by \[\Delta(x^{n+1}+\lambda_1x^{n-1}+\lambda_2x^{n-2}+\dots+\lambda_n)=0.\]
 In the case where $n=2$, \emph{i.e.} when $f\sim x^3$ is a singularity of type $A_2$, then its discriminant is given by $\Delta(x^3+\lambda_1 x+\lambda_2)=0$, \emph{i.e.} the discriminant is a cubic curve defined
 by $-4\lambda_1^3-27\lambda_2^2=0$.
 \end{ex}
 
 The following example will be useful to prove the main result of the next section.
 
 \begin{ex}\rm
  Consider now a singular germ $(f,0)$ of type $D_n$; then $f\sim x^{n-1}+xy^2$. A basis of the local algebra $\mathcal{O}_2\diagup I_{\nabla (x^{n-1}+xy^2)}$ is 
  $(1,x,\dots,x^{n-2},y)$ and, hence, a miniversal deformation is \[F(x,y,\lambda)=x^{n-1}+xy^2+\lambda_1x^{n-2}+\dots \lambda_{n-2}x+\lambda_{n-1}+\lambda_ny.\]
  Its discriminant is given by 
  \begin{equation}\label{D4}
                                    \Delta(x^{n-1}+xy^2+\lambda_1x^{n-2}+\dots \lambda_{n-2}x+\lambda_{n-1}+\lambda_ny)=0.
                                   \end{equation}

 \end{ex}

The following lemma proposes an alternative expression of the discriminant of the $D_n$ singularities.
\begin{lemma}\label{lemD4}
 The discriminant of the miniversal deformation of $f\sim x^{n-1}+xy^2$ is the hypersurface $\Sigma_{D_n}\subset \CC^n$ defined by
 \begin{equation}\label{D4bis}\Delta(\lambda_1,\dots,\lambda_n)=\Delta(t^n+\lambda_1t^{n-1}+\dots+\lambda_{n-1}-(\dfrac{1}{2}\lambda_n)^2)=0.\end{equation}
\end{lemma}

\proof Let us denote by $\Sigma\subset \CC^n$ the locus defined by eq. (\ref{D4bis}). To prove that equations 
(\ref{D4}) and (\ref{D4bis}) are equivalent, we will show that $\Sigma=\Sigma_{D_n}$.

To this end, let us characterize the hypersurfaces $\Sigma$ and $\Sigma_{D_n}$. 
Given the definition of the  discriminant, the expression $\Delta(F(t,\lambda))=0$ means there exists $t_0$ such that 
$F(t_0)=0$ and $\dfrac{\partial F}{\partial t}(t_0)=0$. In other words, $(\lambda_1,\dots,\lambda_n)\in \Sigma$ if, and only if, there 
exists $t_0$ such that \begin{equation}\label{D4bissys}\left\{\begin{array}{ccc}
                                t_0^n+\lambda_1t_0^{n-1}+\dots+\lambda_{n-1}t_0-(\dfrac{1}{2}\lambda_n)^2& = & 0,\\
                                nt_0 ^{n-1}+(n-1)\lambda_1t_0^{n-2}+\dots+\lambda_{n-1} & = & 0.
                               \end{array}\right\}\end{equation}
                               
 Similarly, $(\lambda_1,\dots,\lambda_n)\in \Sigma_{D_n}$ if, and only if, there exists $(x_0,y_0)$ such that $F(x_0,y_0,\lambda)=\dfrac{\partial F}{\partial x}(x_0,y_0,\lambda)=\dfrac{\partial F}{\partial y}(x_0,y_0,\lambda)=0$, \emph{i.e.}
 \begin{equation}\label{D4sys}
  \left\{\begin{array}{ccc}
  x_0^{n-1}+x_0y_0^{2}+\lambda_1x_0 ^{n-2}+\dots +\lambda_{n-2}x_0+\lambda_{n-1}+\lambda_ny_0 &= &0,\\
         (n-1)x_0^{n-2}+y_0^{2}+(n-2)\lambda_1x_0 ^{n-3}+\dots \lambda_{n-2} &= &0,\\
         2x_0y_0+\lambda_n  & = & 0.
         \end{array}\right\}
 \end{equation}

 Let us assume that $\lambda_n\neq 0$, then if $(\lambda_1,\dots,\lambda_n)\in \Sigma$ there exists $t_0$ such that the system (\ref{D4bissys}) is satisfied. It is obvious that $\lambda_n\neq 0$ implies 
 $t_0\neq 0$ and thus one can check that the system (\ref{D4sys}) is also satisfied for $(x_0,y_0)=(t_0,-\dfrac{\lambda_n}{2t_0})$. 
 This proves that $(\lambda_1,\dots,\lambda_n)\in \Sigma_{D_n}$. On the other hand, 
 if $(\lambda_1,\dots,\lambda_n)\in \Sigma_{D_n}$
 and $(x_0,y_0)$ is a solution of (\ref{D4sys}), then necessarily $y_0=-\frac{\lambda_n}{x_0}$. One can further show that $t_0=x_0$ is a solution of (\ref{D4bissys}) and, therefore, 
 $(\lambda_1,\dots,\lambda_n)\in \Sigma$.
 Let us now consider the case $\lambda_n=0$. Then $(\lambda_1,\dots,\lambda_n)\in \Sigma$ for a given $t_0$ implies $(\lambda_1,\dots,\lambda_n)\in \Sigma_{D_n}$
 for $(x_0,y_0)=(t_0,0)$. On the other hand, let us assume $(\lambda_1,\dots,\lambda_n)\in \Sigma_{D_n}$ for a given $(x_0,y_0)$. The equation $2x_0y_0+\lambda_n=0$ forces $x_0$ or $y_0$ to be zero.
 But if $x_0=0$ then necessarily also $a_{n-1}=0$ and $t_0=0$ is a solution of (\ref{D4bissys}), proving $(\lambda_1,\dots,\lambda_n)\in \Sigma$. If $x_0\neq 0$, then $y_0=0$ and
 $t_0=x_0$ is a solution of (\ref{D4bissys}), proving again $(\lambda_1,\dots,\lambda_n)\in \Sigma$. $\Box$

\subsection{Hyperdeterminant of format $2\times 2\times 2\times 2$  and $D_4$-discriminant}\label{hyperdeterminant} 
The  hyperdeterminant of format $2\times 2\times 2\times 2$ is a SLOCC-invariant polynomial 
generalizing the ideas of Cayley for defining a higher dimensional counterpart
of the determinant for 
multimatrices. From a geometrical perspective, the hyperdeterminant 
and its generalization have been studied by 
Gelfand, Kapranov and Zelevinsky\cite{GKZ} in terms of the concept of dual varieties. The geometric definition is the following one: Let $X\subset \PP(V)$ be a (smooth) projective variety, we denote by $X^*$ the dual variety of $X$, defined by
\[X^*=\overline{\{H\in \PP(\mathcal{H}^*), \exists x\in X, T_x X\subset H\}}.\]
For the case $X=\PP^1\times\PP^1\times\PP^1\times\PP^1$, the dual variety, denoted $X^*$, is a 
SLOCC-invariant hypersurface, whose equation is called the hyperdeterminant of format $2\times 2\times 2\times 2$.
This invariant polynomial, denoted as $\Delta_4$, is an irreducible polynomial ($X^*$ is irreducible because $X$ is), 
its degree is $24$, and the corresponding hypersurface is singular\cite{WZ,My2} in codimension $1$.
By definition, $X^*$ parametrizes the singular hyperplane sections of $X$ (alternatively, $H\notin X^*$ is equivalent to saying that $X\cap H$ is a smooth section).

It would be difficult to quote all the papers in QIT (as well as in theoretical physics) 
referring to the concept of hyperdeterminant\cite{BDDER2,BDFMR,HLT,LSV,LT,My,My2}, but it is clear that this invariant 
polynomial plays a central role in understanding the symmetries involved in the SLOCC group action.

In the case of four-qubit systems, the ring of polynomials invariant under the group SLOCC was determined by Luque and Thibon\cite{LT}.
It is a finitely-generated ring with four generators  $B$, $L$, $M$ and $D$, of respective degrees $2,4,4$ and $6$ (explicit expressions, with the same notations, can be found in Ref\cite{HLT2}).
In other words, any SLOCC-invariant polynomial $P$ over $\mathcal{H}=\CC^2\otimes\CC^2\otimes\CC^2\otimes \CC^2$ belongs to $\CC[B,L,M,D]$.
In particular, the hyperdeterminant of format $2\times2\times 2\times 2$ can be expressed as a polynomial in the generators of the ring of invariants and one gets\cite{LT}
\[\Delta_4=\frac{1}{256}(S^3-27T^2),\]

with $S=\dfrac{1}{12}(B^2-4(L+M))^2-24(BD+2LM)$ and $T=\dfrac{1}{216}((B^2-4(L+M))^3-3(B^2-48(L+M))(BD+2LM)+216D^2)$.
In his attempt to give a geometric meaning of the invariants of Luque and Thibon, L\'evay\cite{Levay} introduced some alternatives generators
which are related to the previous ones as
$I_1=\frac{1}{2}B$, $I_2=\dfrac{1}{6}(B^2+2L-4M)$, $I_3=D+\frac{1}{2}BL$ and $I_4=L$. L\'evay's motivation to define this new set of generators was to obtain a more geometrical and 
uniform description of those polynomials, as it is shown in his paper\cite{Levay}. These news invariants $I_1,I_2,I_3,I_4$ allow one 
to get a new expression of $\Delta_4$. In particular, L\'evay proved (Eq (56)\cite{Levay}) that 
\begin{equation}\label{quartic}
\Delta_4=\frac{1}{256}\Delta(t^4-(4I_1)t^3+(6I_2)t^2-(4I_3)t+I_4^2)
\end{equation}
(where $\Delta$ is the discriminant of the polynomial in the $t$ variable).
This particular finding leads to the following claim:

\begin{prop}\label{dual}
 Let us consider the quotient map $\Phi:\mathcal{H}\to \CC^4$ defined by 
\[\Phi(|\Psi\rangle)=(\tilde{I}_1(|\Psi\rangle),\tilde{I}_2(|\Psi\rangle),\tilde{I}_3(|\Psi\rangle),\tilde{I}_4(|\Psi\rangle),\]
where $\tilde{I}_1=-{4}I_1$, $\tilde{I}_2={6}I_2$, $\tilde{I}_3=-{4}I_3$ and $\tilde{I}_4=\dfrac{i}{{2}}I_4$.
Then, $\Phi(\widehat{X}^*)=\Sigma_{D_4}$. 
\end{prop}

\proof According to L\'evay's equation for the hyperdeterminant $\Delta_4$, it is clear that our choice of $\Phi$ 
implies that the equation of $\Phi(\widehat{X}^*)\subset \CC^4$ is 
\[\dfrac{1}{256}\Delta(t^4+\lambda_1t^3+\lambda_2 t^2+\lambda_3t-(\dfrac{1}{2}\lambda_4)^2)=0,\] where $(\lambda_1,\lambda_2,\lambda_3,\lambda_4)$ are coordoninates on $\CC^4$. But Lemma \ref{lemD4} implies that this zero locus is the
discriminant of the $D_4$ simple singularity, \emph{i.e.} the hypersurface $\Sigma_{D_4}$. $\Box$

\begin{rem}\rm
 Propositions \ref{section} and \ref{dual} prove Theorem \ref{mainthm}.
\end{rem}

\begin{rem}\rm
 The quartic $t^4-(4I_1)t^3+(6I_2)t^2-(4I_3)t+I_4^2$ of Eq (\ref{quartic}) appears also
 in the conclusion of a previous paper involving the first two authors\cite{HLT2}. When we evaluate this quartic on the $G_{abcd}$ state, \emph{i.e.} 
 when we consider 
 the quartic $Q(t)=t^4-(4I_1(G_{abcbd}))t^3+(6I_2(G_{abcd}))t^2-(4I_3(G_{abcd}))t+I_4(G_{abcd})^2$, one 
 obtains $Q(t)=(t-a^2)(t-b^2)(t-c^2)(t-d^2)$.
 The state $G_{abcd}$ will cancel $\Delta_4$ if and only if the quartic $Q$ has (at least) a repeated root, \emph{i.e.} there is
 (at least) a relation (among the parameters)
 of type
 $m=\pm n$ with $m\in\{a,b,c,d\}$ and $n\in \{a,b,c,d\}\setminus m$.
 Obviously this condition is satisfied by all values of the parameters $\{a,b,c,d\}$ of Tables \ref{Gabcd} and \ref{Gabcd_nonisole} because
 the corresponding 
 states belong to the dual of $X$ (and thus vanish $\Delta_4$). However the relations between the hyperplane sections of Tables \ref{Gabcd} and \ref{Gabcd_nonisole}
and the number of repeated roots of the quartic $Q$ is probably worth to be further investigate.
\end{rem}

\begin{rem}\rm
 Proposition \ref{dual} establishes a connexion between two types of discriminant. As pointed out earlier, the dual variety of $X$ is 
 a discriminant in the sense that it parametrizes the singular hyperplane sections of $X$. The $D_4$-discriminant parametrizes the singular deformation of 
 the germ $x^{3}+xy$. The most singular deformation of $x^{3}+xy^2+\lambda_1x^2 +\lambda_2 x+\lambda_3+\lambda_4y$ is  obtained for 
 $(\lambda_1,\lambda_2,\lambda_3,\lambda_4)=(0,0,0,0)$ .
 The preimage via the quotient map of $(0,0,0,0)$ is given by the zero-locus of
 (all) invariant polynomials 
 \[\Phi^{-1}(0,0,0,0)=\{|\Psi\rangle, \tilde{I}_1(|\Psi\rangle)=\tilde{I}_2(|\Psi\rangle)=\tilde{I}_3(|\Psi\rangle)=\tilde{I}_4(|\Psi\rangle)=0\}.\]
 This set does not depend on our choice of $\Phi$ and, after projectivization, it corresponds to a 
 well-known variety  $\mathcal{N}\subset\PP(\mathcal{H})$,  the nullcone, which was already invoked to 
 describe the entanglement classes of a four-qubit system\cite{BDD,HLT2}.
 As first pointed out in Ref\cite{BDD}, the nullcone admits a stratification into $9$ distinguished classes of orbits which relate to the $9$ families of Verstraete {\em et al.}'s classification.
To emphasize the connexion with the $D_4$ singular type, let us point out that  
$H=\langle \Psi_2|=\langle0000|+\langle1011|+\langle1101|+\langle1110|$ (the hyperplane of Example \ref{exsectionD4}) 
is a smooth point of $\mathcal{N}$ and this characterizes the hyperplanes of $X$ with a $D_4$-singular point.
This correspondence can diagrammatically be sketched as:
\[\begin{array}{ccc}
 X\cap H\sim D_4 &\longleftrightarrow &H\in \mathcal{N}_{smooth}\subset X^*\\
  &  &\left\downarrow\rule{0cm}{0.5cm}\right. \Phi\\
 x^{3}+xy^2 & \longleftrightarrow&(0,0,0,0)\in \Sigma_{D_4}\subset \CC^4.
\end{array}\]

\end{rem}

\section{Conclusion}
We have introduced a new construction that assigns to any quantum state $|\Psi\rangle$ a complex hypersurface defined 
by the hyperplane section $X\cap \langle \Psi|$ of the set $X$ of all separable states. This hypersurface may have singular points, which can be studied using the theory of singularity. Because the variety of separable states is $G$-homogeneous, this construction is $G$-invariant and two states 
$|\Psi_1\rangle$ and $|\Psi_2\rangle$ which do not define equivalent (singular) hyperplane sections will not be SLOCC equivalent. For four qubits, this construction allowed us to realize 
the singularity of type $D_4$ as a specific hyperplane section and we also proved that no ``higher'' isolated singularities can be obtained by this construction.

The $D_4$ singularity is obtained only when we consider the section $X\cap \langle \Psi|$, where  $|\Psi\rangle$ is a point of an orbit of maximal dimension of 
the nullcone\cite{HLT2} (\emph{i.e.} a smooth point of the nullcone). This is emphasized when we rephrase the notion of Cayley $2\times 2\times 2\times 2$ 
hyperdeterminant, \emph{i.e.} the dual equation of the set of separable states, in terms of  the discriminant of a $D_4$-singularity. The stratification of
the discriminant $\Sigma_{D_4}$ in terms of mutiplicities induces a stratification of the dual variety $X^*$ --- a variety that is of great relevance 
in the study of entanglement of four qubits, as pointed out by  Miyake\cite{My,My2}.

Although the correspondence between four qubits and simple Lie algebra 
of type $D_4$ is now clear from the action of the SLOCC group, the correspondence established in this paper  between four qubits and a simple 
singularity of type $D_4$ is rather surprising and points out to a novel relationship between simple Lie algebra and simple singularity of type $D_4$.

\subsection*{Acknowledgment} 

This work is partially supported by the CNRS grant PEPS-ICQ 2013, project CoGIT (Combinatoire et G\'eom\'etrie pour l'InTrication).

The authors thank Metod Saniga and Jean-Yves Thibon for their questions and fruitful comments on earlier version of this work.


\appendix
\section{Hyperplane sections of type $G_{abcd}$}
In this appendix, we will give the different values of the parameters $a,b,c,d$ of the hyperplanes of type
$G_{abcd}$ which lead either to 
hyperplane sections with only $A_1$ singular points (Table \ref{Gabcd}) or hyperplane sections with non-isolated singularities (Table \ref{Gabcd_nonisole}).
\begin{center}
 \begin{table}[!h]
  
  \begin{tabular}{|c|}
  \hline
 $  \{a = a, b = b, c = a, d = d\}, \{a = a, b = b, c = c, d = b\},$\\
 $\{a = a, b = b, c = c, d = c\}, \{a = a, b = b, c = c, d = -b\},$\\
 $\{a = a, b = b, c = c, d = -c\}, \{a = a, b = b, c = -a, d = d\},$\\
 $\{a = a, b = c, c = c, d = d\}, \{a = a, b = -a, c = c, d = d\},$\\
 $\{a = a, b = -c, c = c, d = d\}, \{a = b, b = b, c = c, d = d\},$\\
 $\{a = c, b = 0, c = c, d = d\}, \{a = d, b = b, c = c, d = d\},$\\
$\{a = -c, b = 0, c = c, d = d\}, \{a = -d, b = b, c = c, d = d\}$\\
\hline
  \end{tabular}

\caption{Hyperplane sections of type $G_{abcd}$ with only $A_1$ singularities.}\label{Gabcd}
 \end{table}
\end{center}

\begin{center}
 \begin{table}[!h]

  \begin{tabular}{|c|}
  \hline
  $ \{a = 0, b = 0, c = 0, d = d\}, \{a = 0, b = 0, c = c, d = 0\},$\\ 
   $\{a = 0, b = b, c = 0, d = 0\}, \{a = a, b = 0, c = 0, d = 0\},$ \\
   $\{a = a, b = d, c = d, d = d\}, \{a = a, b = -c, c = c, d = -c\},$\\ 
   $\{a = a, b = -d, c = d, d = d\}, \{a = a, b = -d, c = -d, d = d\},$\\
   $\{a = b, b = b, c = 0, d = b\}, \{a = b, b = b, c = 0, d = -b\},$\\
   $\{a = c, b = 0, c = c, d = c\}, \{a = c, b = 0, c = c, d = -c\},$\\
   $\{a = c, b = c, c = c, d = d\}, \{a = c, b = -c, c = c, d = d\},$\\ 
   $\{a = d, b = b, c = d, d = d\}, \{a = d, b = d, c = c, d = d\},$\\
   $\{a = d, b = d, c = d, d = d\}, \{a = d, b = -d, c = d, d = d\},$\\
   $\{a = -b, b = b, c = 0, d = b\}, \{a = -b, b = b, c = 0, d = -b\},$\\ 
   $\{a = -b, b = b, c = c, d = -b\}, \{a = -c, b = 0, c = c, d = c\},$\\
   $\{a = -c, b = 0, c = c, d = -c\}, \{a = -c, b = b, c = c, d = -c\},$\\
   $\{a = -c, b = c, c = c, d = d\}, \{a = -c, b = c, c = c, d = -c\},$\\
   $\{a = -c, b = -c, c = c, d = d\}, \{a = -c, b = -c, c = c, d = -c\},$\\
   $\{a = -d, b = b, c = d, d = d\}, \{a = -d, b = b, c = -d, d = d\},$\\
   $\{a = -d, b = d, c = c, d = d\}, \{a = -d, b = d, c = d, d = d\},$\\
   $\{a = -d, b = d, c = -d, d = d\},\{a = -d, b = -d, c = c, d = d\},$\\
   $\{a = -d, b = -d, c = d, d = d\}, \{a = -d, b = -d, c = -d, d = d\}$\\
   \hline
  \end{tabular}

\caption{Hyperplane sections of type $G_{abcd}$ with non-isolated singularities.}\label{Gabcd_nonisole}
 \end{table}
\end{center}
 

\begin{thebibliography}{99}
\bibitem{A} Arnol'd V., ``Normal forms for functions near degenerate critical points, the Weyl groups of A k, D k, E k and Lagrangian singularities.'' Functional Analysis and its applications 6.4 (1972): 254-272.
\bibitem{BDD}  Borsten L.,  Dahanayake D.,  Duff M. J., Marrani A. and Rubens W.,``Four-Qubit Entanglement Classification from String Theory'', Phys. Rev. Lett. {\bf 105}, 100507 (2010). 
\bibitem{BDDER}  Borsten L.,  Dahanayake D.,  Duff M. J., Ebrahim H. and Rubens W., ``Freudenthal triple classification of three-qubit entanglement'', Physical Review A 80, no 3 (2009):032326.
	
\bibitem{BDDER2} Borsten L., Dahanayake D., Duff M. J., Ebrahim H., and Rubens W.. ``Black holes, qubits and octonions.'' Physics Reports 471, no. 3 (2009): 113-219.
\bibitem{BDFMR} Borsten L., Duff M. J., Ferrara S., Marrani A. and Rubens, W. `` Explicit orbit classification of reducible Jordan algebras and Freudenthal triple systems''. arXiv preprint arXiv:1108.0908. (2011).


\bibitem{brody1}  Brody D.C. and Hughston L. P., ``Geometric quantum mechanics'', Journal of Geometry and Physics {\bf 38}, 19-53 (2001).
%
%
%
%
%
%
%
\bibitem{CD} Chterental O. and Djokovic D., ``Normal forms and tensor ranks of pure states of four qubits'', arXiv preprint quant-ph/0612184 (2006).
\bibitem{DGPSS} Decker W., Greuel G.-M., Pfister G., Sch\"onemann SINGULAR 3.1.6: a computer algebra system for polynomial computations. University of Kaiserslautern 2013. \url{http://www.singular.uni-kl.de}.
\bibitem{Du} Durfee A., ``Fifteen characterizations of rational double points and simple critical points.'' L'Enseignement Math\'ematique 15 (1979): 131-163.

%
\bibitem{GKZ} I.M Gelfand M.M Kapranov A.V. Zelevinsky, 
      {\em Discriminants, Resultants and Multidimensional Determinants},
      Birkh\"auser 1994.
%
%
\bibitem{hey}  Heydari H., ``Geometrical Structure of Entangled States and the Secant Variety'', Quantum Information Processing {\bf 7} (1), 3-32 (2008).
%
%
%
%
\bibitem{HLT} Holweck F., Luque J.-G., Thibon J.-Y., ``Geometric descriptions of entangled states by auxiliary varieties'', Journal of Mathematical Physics {\bf 53}, 102203 (2012).

\bibitem{HLT2} Holweck F. Luque J.-G., Thibon J.-Y., ``Entanglement of four qubit systems: a geometrical atlas with polynomial compass I (the finite world)'', arXiv:1306.6816 (2013).
%
%
%
%
%
%
%
%
\bibitem{Kn} Knop F., ``Ein neuer Zusammenhang zwischen einfachen Gruppen und einfachen Singularit\"aten.'' Inventiones mathematicae 90, no. 3 (1987): 579-604.
%
%
%
%
\bibitem{Levay1} L\'evay P. ``STU black holes as four-qubit systems'', Phys. Rev. D {\bf 82}, 026003 (2010).
\bibitem{Levay} L\'evay P.,. ``On the geometry of four-qubit invariants.'' Journal of Physics A: Mathematical and General 39, no. 30 (2006): 9533.

\bibitem{LSV} Levay P., Saniga M. and Vrana, P. ``Three-qubit operators, the split Cayley hexagon of order two, and black holes''. Physical Review D, 78(12), 124022. (2008).

%
%
\bibitem{LT} Luque J.-G. and  Thibon J.Y, ``The polynomial invariants of four qubits'', Phys. Rev. A {\bf 67}, 042303(2003)
%
%

\bibitem{Mi} Milnor J. {\em Singular Points of Complex Hypersurfaces}.(AM-61). No. 61. Princeton University Press, 1968.

 \bibitem{My} Miyake A., ``Classification of multipartite entangled states by multidimensional determinants'', Phys. Rev. A {\bf 67}, 012108 (2003).
%
%
\bibitem{My2} Miyake A. and Verstraete F., ``Multipartite entanglement in $2\times 2\times n$ quatum systems'', Phys. Rev. A {\bf 69}, 012101 (2004).
%
%



\bibitem{PSH} Planat M., Saniga M. and Holweck, F. ``Distinguished three-qubit 'magicity' via automorphisms of the split Cayley hexagon.'' Quantum Information Processing {\bf 12}, 2535-2549 (2013).
\bibitem{SLPP} Saniga, M., and L\'evay, P. ``Mermin's pentagram as an ovoid of PG(3,\,2),'' EPL / Europhysics Letters {\bf 97}, 50006 (2012). 
%
\bibitem{slodowy1} Slodowy P., {\em Simple singularities and simple algebraic groups}. Vol. 815. Berlin: Springer, 1980.

\bibitem{slodowy2} Slodowy P., ``Platonic solids, Kleinian singularities, and Lie groups'', In Algebraic geometry, pp. 102-138. Springer Berlin Heidelberg, 1983.

%
%
%
%
%
 \bibitem{VDMV}  Verstraete F.,  Dehaene F.,  De Moor B. and  Verschelde H., ``Four qubits can be entangled in nine different ways'', Phys. Rev.. A {\bf 65}, 052112 (2002).
%
%


%
 \bibitem{WZ}  Weyman J. and   Zelevinsky A., ``Singularities of
       Hyperdeterminants'', Annales de l'Institut Fourier {\bf{46}}, 
    591-644 (1996).
\bibitem{Wir} Wirthm\"uller K., ``Singularities determined by their discriminants'', Math. Annalen {\bf 252} (1980), 231-245.
%
\end{thebibliography}
\end{document}